# What Information is Necessary and Sufficient to Predict Materials Properties using Machine Learning?

Siyu Isaac Parker Tian[1,2], Aron Walsh[3], Zekun Ren[1,2,†], Qianxiao Li[4], Tonio Buonassisi[1,5,*]

[1] Low Energy Electronic Systems (LEES), Singapore-MIT Alliance for Research and Technology (SMART), 1 Create Way, Singapore 138602, Singapore

[2] Solar Energy Research Institute of Singapore (SERIS), National University of Singapore, 7 Engineering Drive, Singapore 117574, Singapore

[3] Department of Materials, Imperial College London, Prince Consort Rd, South Kensington, London SW7, United Kingdom

[4] Department of Mathematics, National University of Singapore (NUS), 21 Lower Kent Ridge Rd, Singapore 119077, Singapore

[5] Department of Mechanical Engineering, Massachusetts Institute of Technology (MIT), 77 Massachusetts Ave., Cambridge, MA 02139, USA

[†] Now at: Xinterra, Singapore, 77 Robinson Road, Singapore 068896, Singapore

[*] Corresponding author: T.B. (buonassi@mit.edu)

## Abstract:

Conventional wisdom of materials modelling stipulates that both chemical composition and crystal structure are integral in the prediction of physical properties. However, recent developments challenge this by reporting accurate property-prediction machine learning (ML) frameworks using composition alone *without* knowledge of the local atomic environments or long-range order. To probe this behavior, we conduct a systematic comparison of supervised ML models built on composition only *vs.* composition plus structure features. Similar performance for property prediction is found using both models for compounds close to the thermodynamic convex hull. We hypothesize that composition embeds structural information of ground-state structures in support of composition-centric models for property prediction and inverse design of stable compounds.

## Introduction

What information of a crystalline material is necessary to predict its properties? The word “structure” quickly comes to mind as structure-property relationships are a cornerstone of materials science. Crystal structure, which can be classified into the 7 crystal systems and 230 space groups, is a prominent feature for materials modelling. In atomistic descriptions, the total crystal potential is introduced as an expansion from two-body to many-body interatomic interactions [1]. In the widely applied density functional theory (DFT) formalism, the spatial coordinates of atoms influence the quantum mechanical electron density, and thus all ground-state properties of the material, through the external potential [2], [3].

Surprisingly, a direct link between composition and properties has been established in a series of recent studies for material properties as varied as superconductivity critical temperature [4], hardness [5], and band gap [6]. These statistical machine learning (ML) models are built from features containing only compositional information (elements and stoichiometries) of the materials. Unlike first principles modelling approaches that require no system specific parametrization, the ML models require large datasets for training and testing, typically hundreds to tens of thousands of compounds.

These developments in the interplay between materials composition, structure, and properties invites us to assess whether current materials representations are optimal from the perspectives of scientific interpretability and model accuracy. Based on the evidence presented below, we argue that explicit structural information is not a necessary input to accurately predict many properties using ML for many datasets of interest. The compositional information, able to embed the chemical bonding and connectivity implicitly, is sufficient for constructing accurate and predictive models.

## Materials information in ML frameworks

The evolution of ML frameworks can be considered through the lens of available tools and data. The largest datasets in materials science have traditionally focused on crystal structure information determined from diffraction experiments and reported in the form of crystallographic information files (CIFs). Over the past decade, these have been used to construct systematic databases with predicted structure *and* property information [7]–[10]. However, datasets are often incomplete for more sophisticated properties such as charge transport. For instance, SuperCon [11] contains superconductivity critical temperatures for a series of compositions but with no solved crystal structures. It has been shown that supervised ML models can predict critical temperatures from compositional information alone [4]. Here a minimalist materials representation was necessary due to the absence of structural information.

The trend in machine learning materials has followed an established pattern. When training data is scarce, *e.g.*, at the beginning of a field's exploration, classical ML models with complex, handcrafted representations [12]–[14] are common [15]–[19] (see Figure 1 for examples). Later, as data becomes more available in structured databases, there is a shift from better representations to larger model capacities [20]–[26]. Deep learning models (many layers of neural networks), with better approximating power, automatically learn representations from simple inputs (*e.g.*, chemical formulas). Typically, the deep learning models require either composition only (*Comp*) or composition plus structure (*CompStruct*) inputs, providing a clear distinction between compositional and structural information used. This distinction offers an opportunity to study how ML frameworks interpret materials information embedded in simple inputs to predict materials properties [27], and how these differ from first-principles modelling approaches.

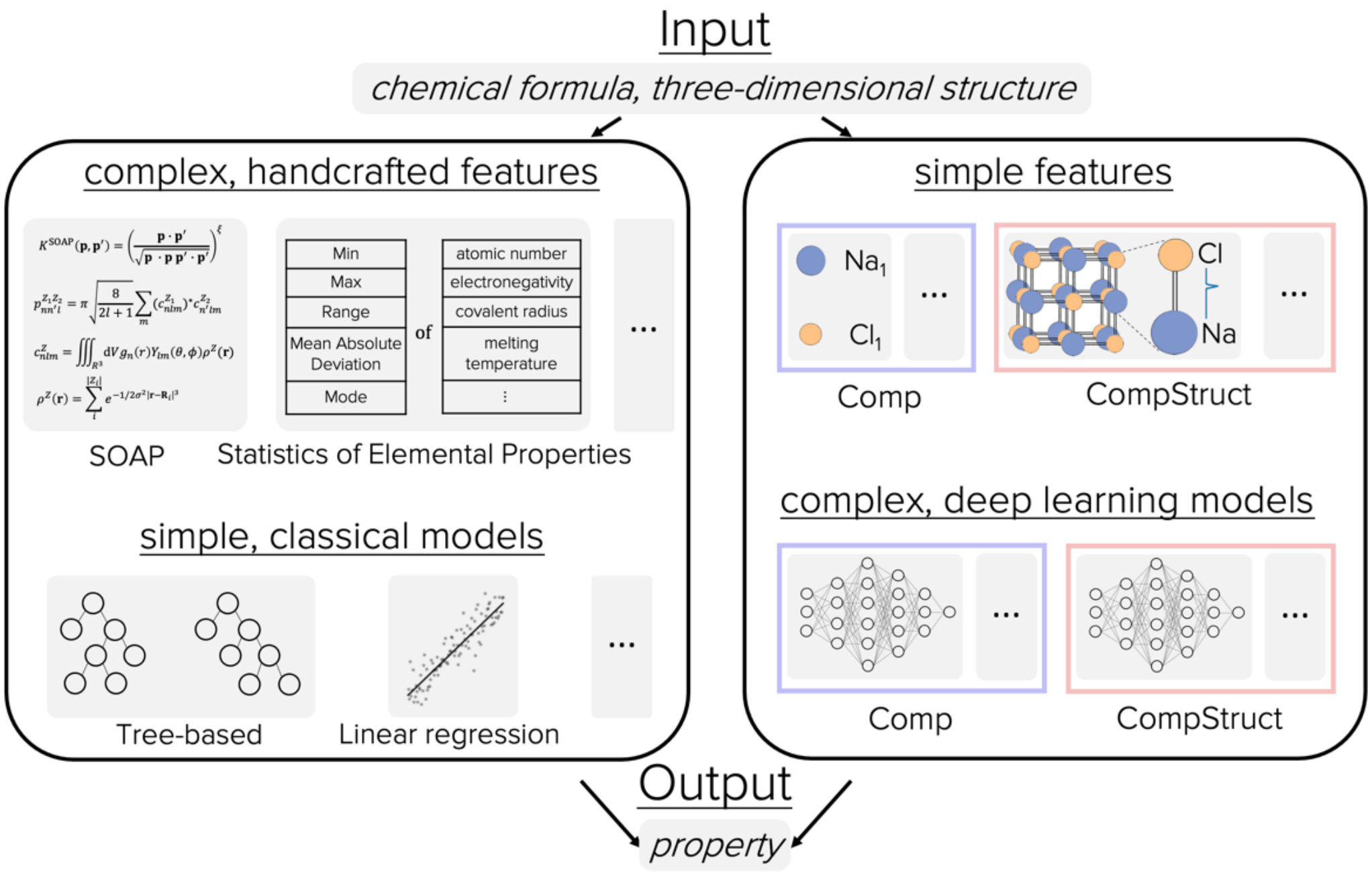

Figure 1. Two workflows for property prediction using machine learning. Composition-structure-property relationships are complex. Historically, machine-learning models embed this complexity predominantly in the features (left), or in the model parameters (right). The latter do not rely on pre-determined assumptions about property-structure-composition relationships but require more training data to yield accurate predictions. The latter typically process either composition only (Comp) or composition plus structure (CompStruct) inputs.

Here, we compare two classes of ML models in respective frameworks: (1) composition only (*Comp*) models, which process only compositional information, *i.e.*, constituent elements and stoichiometry, and (2) composition plus structure (*CompStruct*) models, processing additional structural information, *e.g.*, atom positions and lattice vectors. Our aim is to answer the question: *what is the necessary and sufficient (minimum) materials information to predict a wide range of material properties?* The answer has implications on our understanding of materials in general, and specific implications on the design of inverse-design algorithms to create new forms of matter from user-specified target properties.

## Methodology

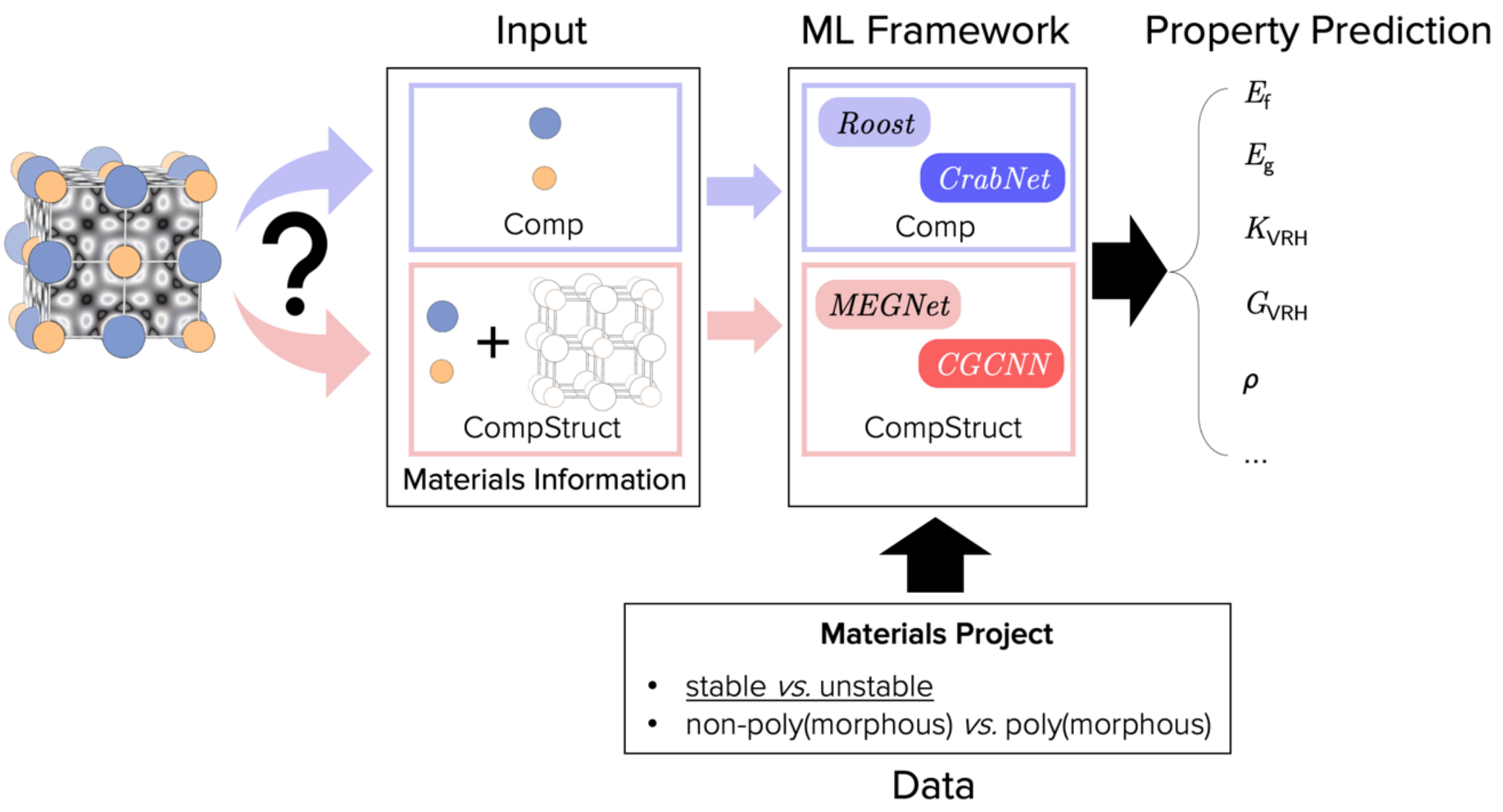

Figure 2. Methodology of comparing composition only (Comp) *vs.* composition plus structure (CompStruct) ML frameworks. Selected frameworks are Roost [22] and CrabNet [25] for Comp frameworks, and MEGNet [21] and CGCNN [20] for CompStruct frameworks. Data are from the Materials Project [7] with a data segregation of stable *vs.* unstable compounds. The properties predicted are formation energy (per atom), $E_{\mathrm{f}}$, band gap, $E_{\mathrm{g}}$, (Voigt-Reuss-Hill average) bulk modulus, $K_{\mathrm{VRH}}$, (Voigt-Reuss-Hill average) shear modulus, $G_{\mathrm{VRH}}$, density, $\rho$, etc.

The overall methodology is shown in Figure 2. To minimize insufficient use of materials information manifested as model inaccuracies, we choose to compare frameworks with state-of-the-art performance in Matbench [15]: Roost and CrabNet for Comp, and MEGNet and CGCNN for CompStruct. To alleviate unfairness of varied amount of elemental information present in individual inputs, we standardize the elemental embedding to one-hot-encoded atomic numbers. The original CGCNN uses one-hot-encoded elemental properties for elemental embedding, while original Roost and CrabNet use natural-language-processing (NLP)-based Mat2Vec [28] elemental embedding. Since MEGNet already uses the atomic number as a categorical identifier to automatically learn the elemental embedding with an embedding layer [29], we retain the original implementation.

For our datasets, we select properties that are widely benchmarked, representative, and of a sizable number (to avoid non-optimal model performance due to data scarcity). Consequently, we select formation energy, $E_{\mathrm{f}}$, band gap, $E_{\mathrm{g}}$, (Voigt-Reuss-Hill average) bulk modulus, $K_{\mathrm{VRH}}$, (Voigt-Reuss-Hill average) shear modulus, $G_{\mathrm{VRH}}$, and density, $\rho$. In addition, to offer a solely structure-dependent property as a baseline, we use another toy property, point density (by substituting atomic mass with the number of atoms in the unit cell in the calculation of density).

To understand the apparent conflict between the good performance of existing Comp models and the physical dictation that structure is indispensable in the prediction of materials properties, we postulate that the conflict arises from the choice of datasets—the good performance of Comp models is usually achieved on datasets comprising only ground-state compounds, while the physical significance of structure is universally established regardless of datasets. Consequently, we investigate the effect of data segregation by expanding and stress testing the status quo to understand the role of the dataset choice in the matter of minimum materials information needed for ML property prediction. There are two dimensions to consider: (i) thermodynamic stability from the materials perspective (ground-state compounds *vs*. including all compounds) [30]; (ii) one-output-property-for-one-input-compound from the data perspective (ground-state compounds *vs.* including all high-energy polymorphs). We thus stress test based on the two matching dimensions and develop the data segregations in this study:

1. *stable vs. unstable* (stress testing stability): below and above 100 meV/atom above the thermodynamic convex hull in chemical potential space, respectively. While informed by heuristics, we acknowledge that other stability metrics have been proposed [31]–[35]. Our choice of a fixed cut-off should include a majority of properly classified compounds that are of high interest.

2. *non-poly vs. poly* (stress testing one-property-value-for-one-composition): non-polymorphous and polymorphous entries have only one structure and many structures, respectively. Note that poly contains ground-state compounds if they have polymorphs. This segregation guarantees one output for one input in terms of Comp frameworks. This is not an issue for CompStruct frameworks because of their inclusion of structural materials information.

For training, we follow the Matbench convention for MEGNet and CGCNN with 60%, 20%, 20% split of training, validation, and test set. To ensure the generality of the results, we average three data splits (training-validation-test splits) with three model initializations, yielding nine runs for every dataset. The dataset refers to each of the data segregation of every property, *e.g.*, stable data of $E_\mathrm{f}$. CrabNet is an exception because it has a specified model initialization seed; thus, we only have three runs of data splits for CrabNet datasets. We report the performance on the test set with a widely reported and benchmarked metric, mean absolute error (MAE) between the predicted property and the actual property.

## Results

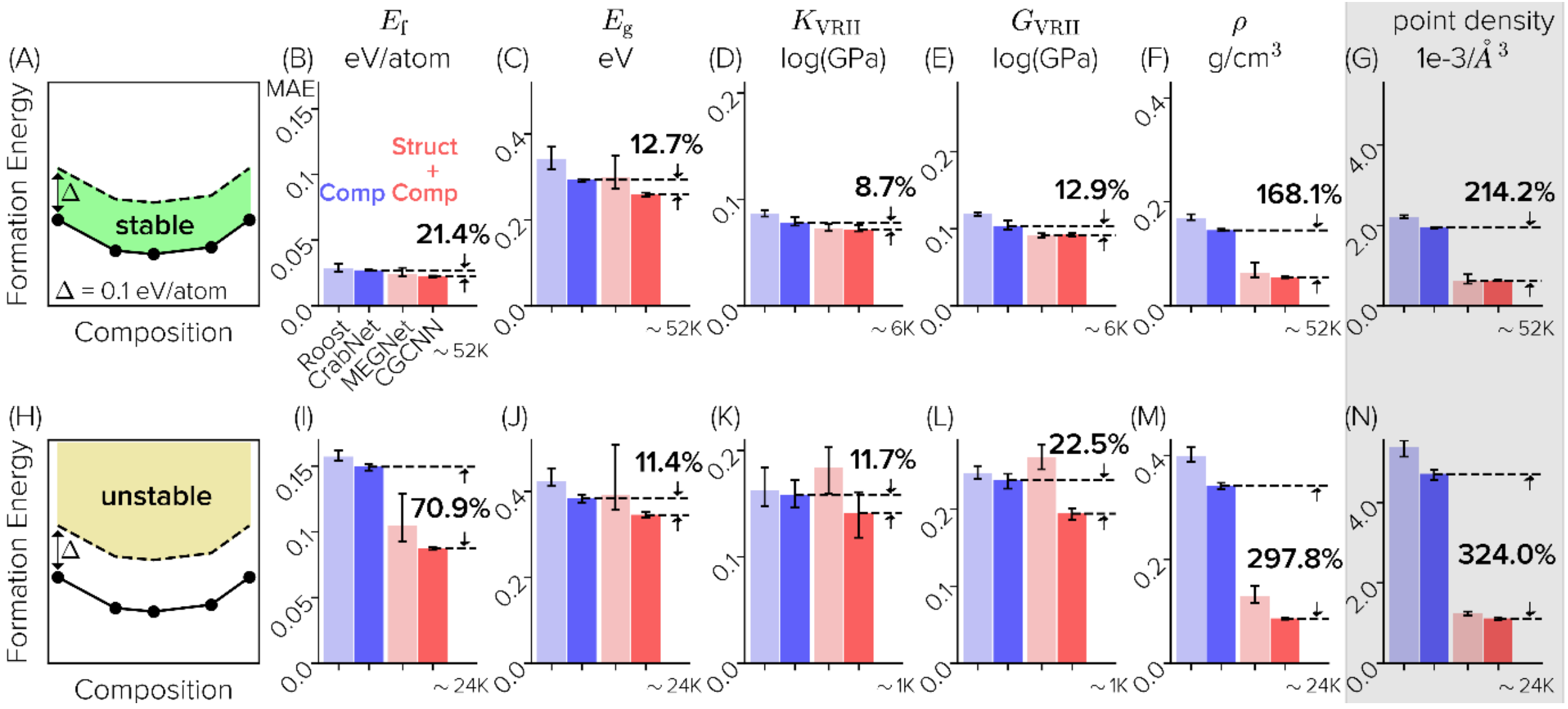

Figure 3. Comp *vs.* CompStruct model performance on stable *vs.* unstable data segregation across properties. (A and H) illustration of the stable *vs.* unstable data segregation. (B and I) The mean absolute

error (MAE) of Comp frameworks (colored in blues), Roost and CrabNet, *vs.* CompStruct frameworks (colored in reds), MEGNet and CGCNN, on the prediction of formation energy, $E_{\mathrm{f}}$, on stable *vs.* unstable data segregation. The bar represents the average MAEs of nine runs and the error bar records the min and max MAEs. The percentages are performance (MAE) deviation of the two better MAEs in each of Comp and CompStruct categories expressed in percentage with respect to the smaller MAE. The approximate number of training data (rounded to the nearest thousand) is recorded on the bottom right of each figure panel. The following are respectively the performance of Comp *vs.* CompStruct frameworks on stable *vs.* unstable data segregation on the prediction of (C and J) band gap, $E_{\mathrm{g}}$ (D and K) Voigt-Reuss-Hill average bulk modulus, $K_{\mathrm{VRH}}$ (E and L) Voigt-Reuss-Hill average shear modulus, $G_{\mathrm{VRH}}$ (F and M) density, $\rho$, and (G and N) point density. Point density is a toy property (shaded in gray) solely dependent on structure, modified from density by substituting the atomic mass with the number of atoms in the unit cell in the density calculation.

The Comp frameworks perform on par with CompStruct ones for stable compounds across properties except for density and point density. Figure 3 plots the results as bar plots indicating the MAE between predicted properties and actual properties. The bar indicates the average MAE of different runs, while the error bar the min and max MAE. Comp results are in blues, while CompStruct ones in reds. The percentage performance deviation between Comp and CompStruct, is calculated as $\frac{|\text{best Comp MAE} - \text{best CompStruct MAE}|}{\min(\text{best Comp MAE}, \ \text{best CompStruct MAE})} \times 100\%$. The data segregation, stable *vs.* unstable, manifests as upper row vs. lower row, while the first columns illustrate the respective segregations. The following columns record the results for various properties, including the structure-dependent toy property, point density, shaded in gray. The performance deviation stays small ($\leq \sim 20\%$) for the stable segregation across properties except for density and point density, and many properties exhibit larger performance deviations for the unstable segregation. The small performance deviations are with respect to (w.r.t.) the MAE and tend to shrink further if considered in a relative fashion, *e.g.,* performance deviation w.r.t. the mean absolute percentage error (although not applicable to zero or near-zero property values). This small performance deviation between Comp and CompStruct frameworks for the stable segregation suggests *compositional information alone for stable compounds is sufficient for the prediction of many properties*.

We exclude the attribution of close Comp-CompStruct performances to one-property-value-for-one-composition by assessing non-poly *vs.* poly results in Figure S1. The performance deviation between Comp and CompStruct is generally smaller for the non-poly segregation but still considerably larger than the performance deviation for the stable segregation except for the elastic moduli. This observation suggests one-property-value-for-one-composition is a contributing factor to close Comp-CompStruct performance, but not dominating over stability. The exceptions of elasticity can be explained by their almost all-stable data compositions in the non-poly segregation: two contributing factors of small performance deviations, namely being stable and non-poly, are almost both satisfied, resulting in small performance deviations.

After establishing that Comp models perform on par with CompStruct in the stable segregation, we further strengthen the indicative link to our central hypothesis. First, we exclude insufficient use of structural information by CompStruct models—CompStruct models exhibit considerably lower MAEs than Comp ones for the structure-dependent toy property, point density, even in the stable segregation. Second, we exclude data scarcity. CompStruct models generally possess more parameters, and thus are more susceptible to data scarcity; the CompStruct models yield small MAEs and have small differences between training and test MAEs (Table S1) for the stable segregation, indicating little overfitting, and thus data sufficiency.

Additional test runs using a half and a third of the data show similar performance deviation trends in stable *vs.* unstable segregation. The exception of density to our conclusion shows the limits for an explicitly structure-dependent property, as density is inversely proportional to the unit cell volume.

After demonstrating the sufficiency of compositional information, we verify the auxiliary aspect of our question of the minimum materials information, *i.e.*, the necessity of compositional information, by comparing CompStruct models with dummy Comp inputs (treating all input atoms as hydrogen) with normal CompStruct models in Figure S2 and S3. The dummy Comp input is a substitute for property prediction models taking only structural information (Struct models) due to the lack of the application to predict properties from structure alone. We observe, especially in comparison with the performance deviation between Comp and CompStruct models, structural information alone does not predict properties with enough accuracy evidenced by the large performance deviations in Figure S2 and S3, suggesting *compositional information is necessary*. Although this comparison of models with their partial selves serves as only circumstantial evidence, the small performance deviations of point density throughout data segregations strengthen the inference of the necessity of compositional information by reflecting the non-necessity of compositional information in the case of the structure-dependent toy property of point density.

## Discussion

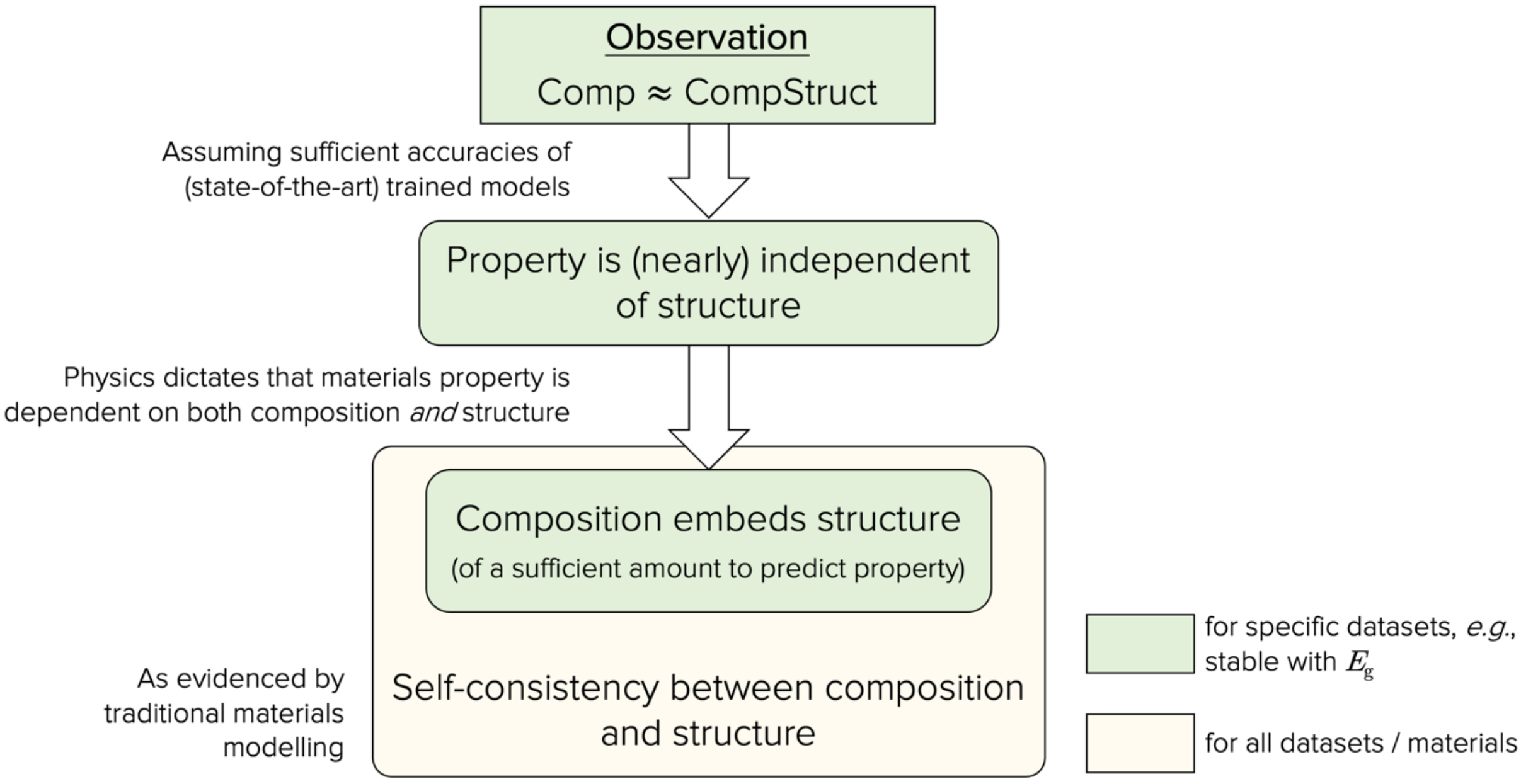

Figure 4. Chain of implications and hypotheses for composition-structure-property relationships in materials.

Our results challenge us to acknowledge that current ML frameworks for specific datasets, *e.g.*, stable with $E_g$, seemingly do not require structural materials information to make property predictions that rival traditional materials modelling in terms of accuracy and precision. To reconcile the apparent violation of physics, where materials property is dependent on both composition and structure, we proffer the following chain of implications and hypotheses as shown in Figure 4:

- From the observation that the performance of Comp frameworks is on par with the performance of CompStruct ones for specific datasets, *i.e.*, stable segregation with properties $E_f$, $E_g$, $K_{VRH}$, and $G_{VRH}$, we imply that these properties are nearly independent of structure within the said datasets, assuming sufficient accuracies of state-of-the-art trained models. (Reason: the statistical independence of these properties on structure conditional on the given datasets is observed from little gains of property prediction performances with the addition of structural information.)
- From the independence of the properties on structure in these datasets, we imply that composition must embed structure of a sufficient amount to predict these said properties. (Reason: the apparent independence on structure must be compensated by the implicit embedding of structure in composition since physics dictates that property is dependent on both composition and structure.)
- We hypothesize the embedding of structure in composition results from a natural self-consistency between composition and structure given specific datasets. Without the constraints of certain datasets, *i.e.*, for all crystalline materials, high diversity of local atomic environments and connectivity occurs, albeit a universal underlying self-consistency between composition and structure to reach local minima of energy configurations. Within specific datasets, *e.g.*, stable segregation with $E_g$, the corresponding structures per composition become restricted, and structure can be (nearly) represented as a function of composition through self-consistency. In the extreme case of ground-state compounds only, there exists a single ground-state structure that will satisfy the local and long-range chemical bonding preferences of the component elements of a given composition (a one-to-one correlation between composition and structure, namely structure as a function of composition) [36]. A perspective on these relations through the lens of implicit function is explored in Section *S3*.

## Implications

The intimate relationship between composition and structure highlighted here widens the application domain of models built using only compositional features. One is for the development of structure prediction tools by exploiting the structural information that is embedded in compositional representations and the high correlation between composition and structure. A second implication is for property prediction when stable compounds are of primary interest, where the use of compositional frameworks that are less data-intensive and quicker to train without much performance loss. A third is toward the design of stable compounds, where the inverse design of candidate composition could be targeted, which avoids the complexity of navigating the higher dimensional structural space. In addition to requiring less data and faster training, inversely designing compositions also adds advantages for synthesis due to the assumption of stability in the designed compositions. Of course, there will be cases where structure knowledge is necessary or beneficial such as the exploration of metastable phases using non-equilibrium conditions (*e.g.*, electrochemical deposition with large bias voltage, or high-energy vapor-deposition methods) and structure features should be considered during inverse design.

## Conclusions

In summary, we compare property predictions for materials using machine learning models with knowledge of structural and/or compositional features for a diverse dataset. We observe that the crystal structure information provides no statistical advantage, except for a property (density) that is explicitly structure-dependent for datasets comprising stable compounds. We conclude that *compositional information alone for stable compounds is sufficient and necessary for the prediction of many properties*. To reconcile with the traditional view of materials modelling that three-dimensional structure is essential, we suggest a revised view on the structure-property relationship: self-consistency between composition and structure is universal, but given certain datasets, *e.g.*, stable compounds, the composition can embed the innate structural preferences of elements with sufficient complexity for property prediction. The implications are two-fold: (1) scientific— advancing our understanding of crystalline materials and the development of structure-prediction models; (2) practical—promoting the priority of using compositional models for property prediction and suggesting for inverse design designing composition alone might be sufficient to identify stable compounds.

## Detailed Methods

For the use of one-hot-encoded atomic numbers as the standard elemental embedding, we inherit the provided one-hot encodings from Roost and CrabNet, and we build our custom one-hot encodings of atomic numbers in the case of CGCNN where there is none provided— atomic number of 94, the last element present in our datasets, is taken as the last element included in the embedding, namely the one-hot encodings have 94 dimensions. The 94 dimensions of the custom-built one-hot encoding is close to the 93 dimensions of the original elemental embedding. For data preprocessing, we adopt the procedures for each property as used in Matbench with necessary modifications, *e.g.*, we didn't implement removing $E_{\mathrm{f}}$ or energy above the convex hull more than 150 meV/atom because we are also interested in the unstable regime of the data. We list our prepressing procedures for each property datasets in Table 1.

Table 1. Preprocessing procedures for various properties; the preprocessing procedures of the toy property, point density, follow the ones of density.

| Property [unit] | Preprocessing |
|---|---|
| $E_{\mathrm{f}}$ [eV/atom] | 1, 2 |
| $E_{\mathrm{g}}$ [eV] | 2 |
| $K_{\mathrm{VRH}}$ [log(GPa)] | 2, 3, 4 |
| $G_{\mathrm{VRH}}$ [log(GPa)] | 2, 3, 4 |
| $\rho$ [g/cm$^3$] | 2 |
| point density [1e-3/Å$^3$] | 2 |

1. remove data whose $E_{\mathrm{f}} > 3$ eV/atom

2. remove data containing noble gases.

3. remove data whose $G_{\mathrm{Voigt}}$, $G_{\mathrm{Reuss}}$, $G_{\mathrm{VRH}}$, $K_{\mathrm{Voigt}}$, $K_{\mathrm{Reuss}}$, or $K_{\mathrm{VRH}} \leq 0$, where $G_{\mathrm{Voigt}}$, $G_{\mathrm{Reuss}}$, $K_{\mathrm{Voigt}}$, and $K_{\mathrm{Reuss}}$ are respectively Voigt average shear modulus, Reuss average shear modulus, Voigt average bulk modulus, and Reuss average bulk modulus.

4. remove data which fail $G_{\mathrm{Voigt}} \leq G_{\mathrm{VRH}} \leq G_{\mathrm{Voigt}}$, or $K_{\mathrm{Voigt}} \leq K_{\mathrm{VRH}} \leq K_{\mathrm{Voigt}}$

## Acknowledgements

We appreciate helpful discussions with James Kirkpatrick (DeepMind), Taylor Sparks (U. of Utah), Anthony Yu-Tung Wang (Technische Universität Berlin), Jakoah Brgoch (U. of Houston), Alexandra Carvalho (National U. of Singapore), Kedar Hippalgaonkar (A*STAR, Nanyang Technological U.), Jiali Li (National U. of Singapore), Ruiming Zhu (A*STAR, Nanyang Technological U.), Maung Thway (A*STAR, Nanyang Technological U.). This research is supported by the National Research Foundation, Prime Minister's Office, Singapore under its Campus for Research Excellence and Technological Enterprise (CREATE) program through the Singapore Massachusetts Institute of Technology (MIT) Alliance for Research and Technology's Low Energy Electronic Systems (LEES) research program. We acknowledge the MIT SuperCloud and Lincoln Laboratory Supercomputing Center for providing high-performance computing and consultation resources that have contributed to the research results reported within this study. We appreciate helpful feedbacks to our arXiv preprint from Shyue Ping Ong, Prashun Gorai, Sterling Baird, Taylor Sparks, Michael Webb *et al.* on stability, observations of similar trends in other materials, and reference suggestions.

## Data and Code Availability

The datasets and code used in this study are all available in the respective following figshare and GitHub repositories: https://figshare.com/articles/dataset/data_tar_gz/20161235 and https://github.com/PV-Lab/CompStruct.

## Author Contributions

S.I.P.T. and T.B. conceived of and designed the study. A.W. and Q.L. provided guidance on density-functional theory, materials representations, and machine-learning methods. S.I.P.T. and Z.R. performed the ML modelling. S.I.P.T. and T.B. wrote the paper, while all co-authors vigorously engaged in framing the arguments in and reviewing the manuscript.

## Conflicts of Interest (COI)

Although our laboratory has IP filed covering photovoltaic technologies and materials informatics broadly, we do not envision a direct COI with this study, the content of which is open sourced. Two of the authors (Z.R. and T.B.) own equity in Xinterra Pte Ltd, which applies machine learning to accelerate novel materials development.

# Supplemental Information

## S1. Comp *vs.* CompStruct results for non-poly *vs.* poly data segregations

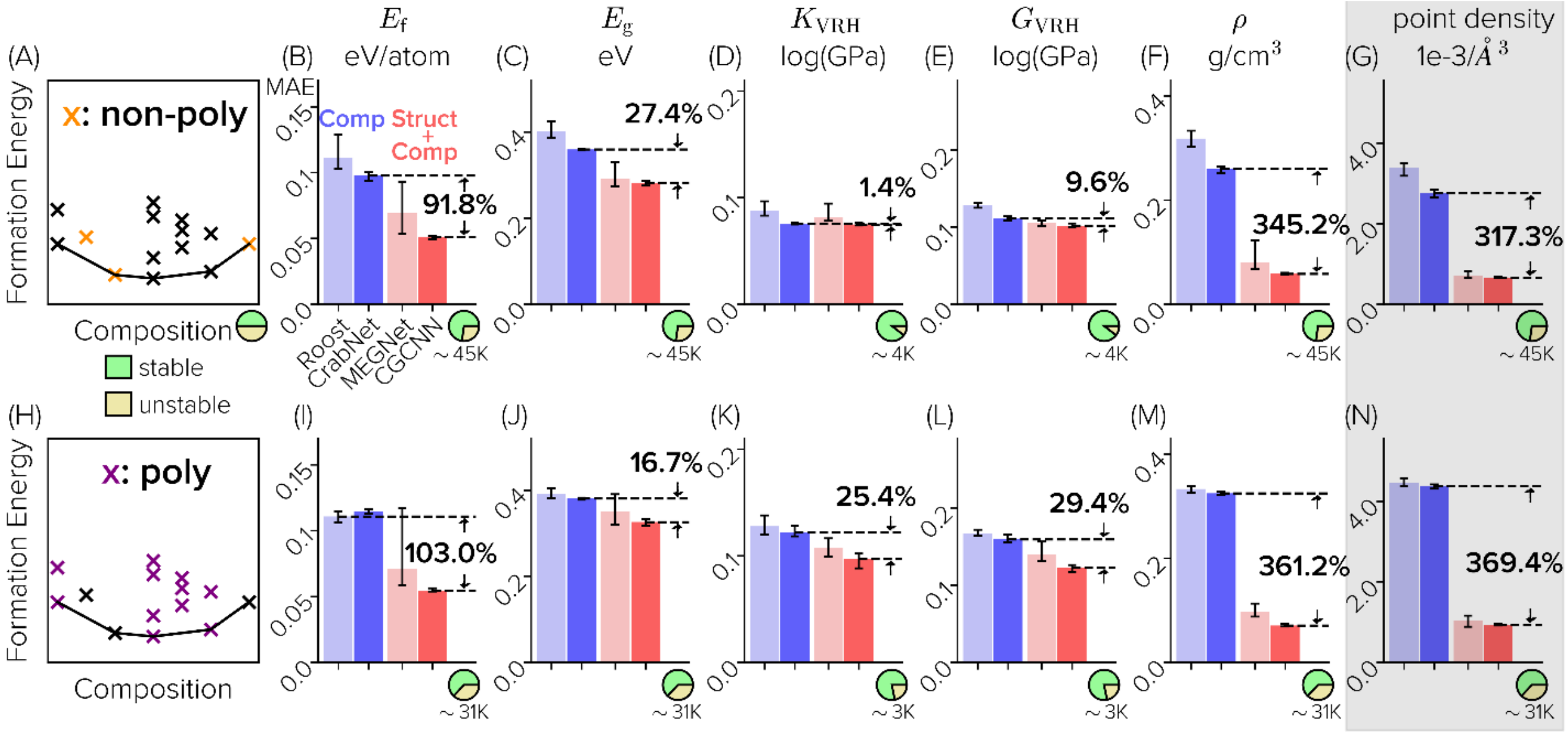

Figure S1. Comp *vs.* CompStruct framework performances on non-poly *vs.* poly data segregation across properties. (A and H) illustration of the non-poly(morphous) *vs.* poly(morphous) data segregation. (B and I) The mean absolute error (MAE) of Comp frameworks (colored in blues), Roost and CrabNet, *vs.* CompStruct frameworks (colored in reds), MEGNet and CGCNN, on the prediction of formation energy, $E_{\mathrm{f}}$, on stable *vs.* unstable data segregation. The bar represents the average MAEs of nine runs over different data splits and model initializations, and the error bar records the min and max MAEs of different runs. The percentages are performance (MAE) deviation of the two better MAEs in each of Comp and CompStruct categories expressed in percentage with respect to the smaller MAE. The pie chart underneath shows the ratio of stable (green) / unstable (yellow) in respective datasets. (The definition of stable / unstable can be found in Figure 3 and S2.) The approximate number of training data (rounded to the nearest thousand) is recorded on the bottom right of each figure panel. The following are respectively the performance of Comp *vs.* CompStruct frameworks on stable *vs.* unstable data segregation on the prediction of (C and J) band gap, $E_{\mathrm{g}}$ (D and K) Voigt-Reuss-Hill average bulk modulus, $K_{\mathrm{VRH}}$ (E and L) Voigt-Reuss-Hill average shear modulus, $G_{\mathrm{VRH}}$ (F and M) density, $\rho$, and (G and N) point density. Point density is a toy property (shaded in gray) solely dependent on structure, modified from density by substituting the atomic mass with the number of atoms in the unit cell in the density calculation.

## S2. CompStruct with dummy Comp inputs *vs.* CompStruct results

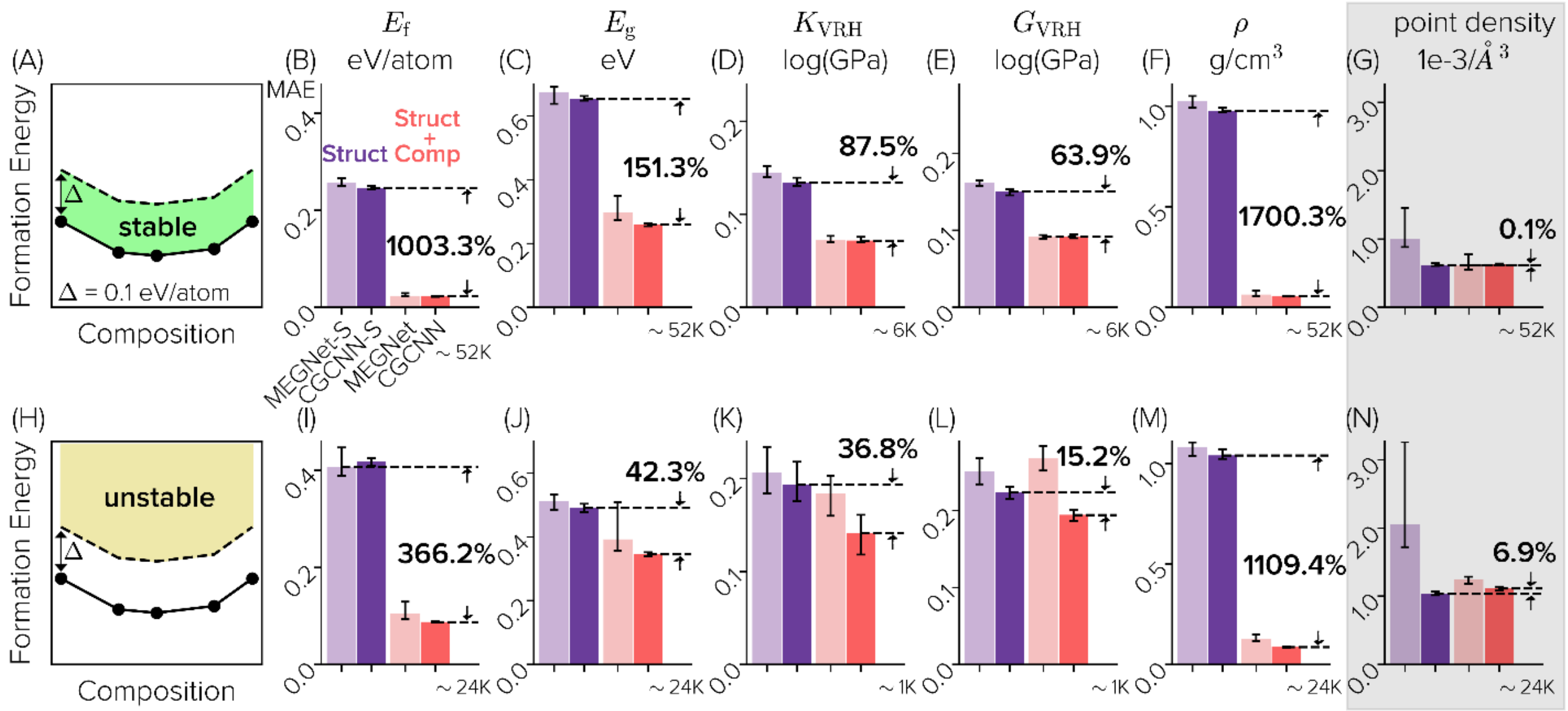

Figure S2. CompStruct framework with dummy Comp inputs (Struct) *vs.* CompStruct (Comp + Struct) framework performances on stable *vs.* unstable data segregation across properties. Another way to evaluate the role of structure, is to replace the "Comp" information with hydrogen. We call this "CompStruct with dummy Comp," and compare it to CompStruct with original composition information. CompStruct frameworks with dummy Comp inputs are colored in purple (the suffix "-S" indicates structural inputs only); CompStruct with original composition information in red. (A and H) illustration of the stable *vs.* unstable data segregation. (B and I) The mean absolute error (MAE) of CompStruct frameworks with dummy Comp inputs, MEGNet-S and CGCNN-S *vs.* CompStruct frameworks, MEGNet and CGCNN, on the prediction of formation energy, $E_f$, on stable *vs.* unstable data segregation. The bar represents the average MAEs of nine runs over different data splits and model initializations, and the error bar records the min and max MAEs of different runs. The percentages are performance (MAE) deviation of the two better MAEs in each of structure-only CompStruct and CompStruct categories expressed in percentage with respect to the smaller MAE. The approximate number of training data (rounded to the nearest thousand) is recorded on the bottom right of each figure panel. The following are respectively the performance of structure-only CompStruct *vs.* CompStruct frameworks on stable *vs.* unstable data segregation on the prediction of (C and J) band gap, $E_g$ (D and K) Voigt-Reuss-Hill average bulk modulus, $K_{VRH}$ (E and L) Voigt-Reuss-Hill average shear modulus, $G_{VRH}$ (F and M) density, $\rho$, and (G and N) point density. Point density is a toy property (shaded in gray) solely dependent on structure, modified from density by substituting the atomic mass with the number of atoms in the unit cell in the density calculation.

In the results shown in Figure S2, one out of nine runs of stable segregation with point density for MEGNet-S didn't train (where the seed for data split is 20, and the seed for model initialization is 1), and the corresponding result is removed (not shown) as an outlier.

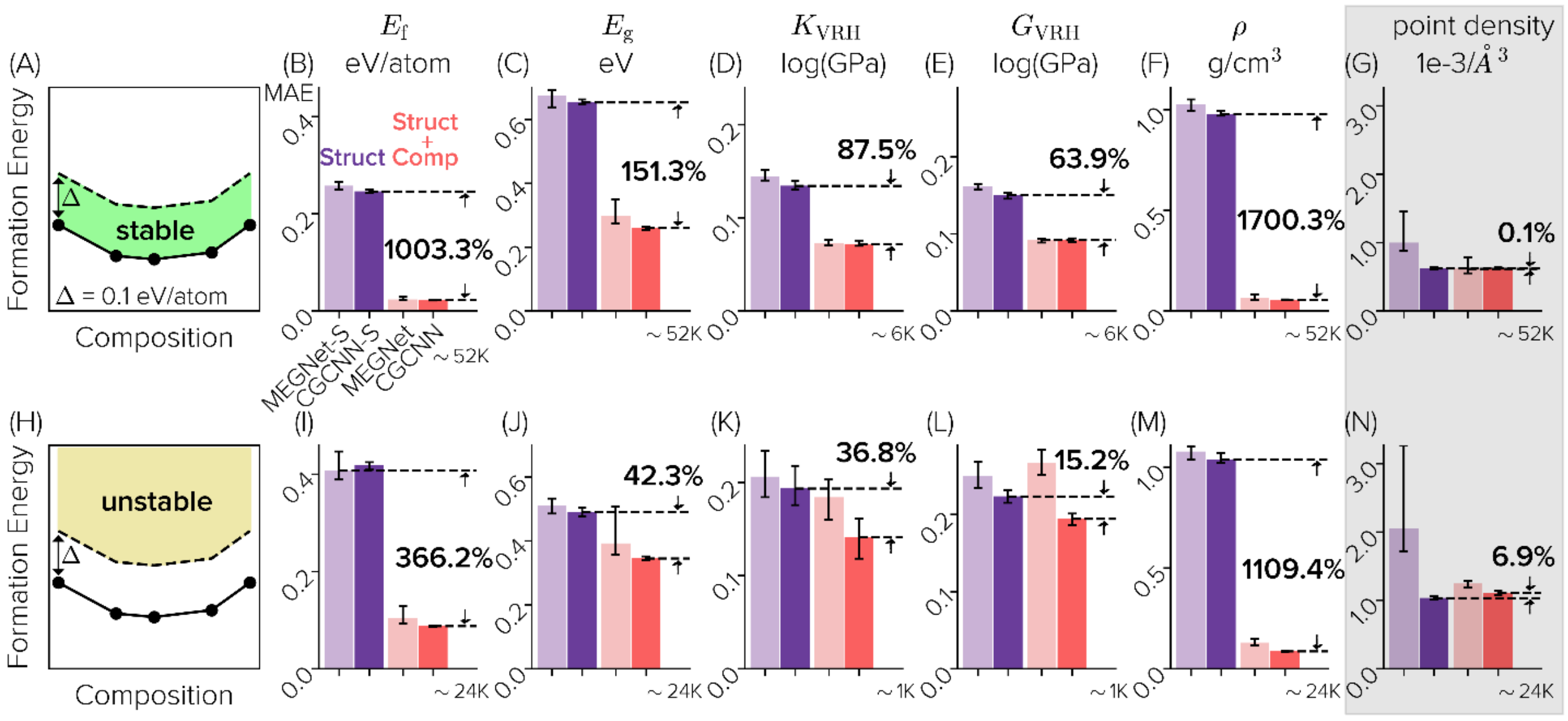

Figure S3. CompStruct framework with dummy Comp (Struct) inputs *vs.* CompStruct (Comp + Struct) framework performances on non-poly *vs.* poly data segregation across properties. Labels are consistent with those described in Figure S2 caption. (A and H) illustration of the non-poly(morphous) *vs.* poly(morphous) data segregation. (B and I) The mean absolute error (MAE) of CompStruct frameworks with dummy Comp inputs, MEGNet-S and CGCNN-S *vs.* CompStruct frameworks, MEGNet and CGCNN, on the prediction of formation energy, $E_\mathrm{f}$, on non-poly *vs.* poly data segregation. The bar represents the average MAEs of nine runs over different data splits and model initializations, and the error bar records the min and max MAEs of different runs. The percentages are performance (MAE) deviation of the two better MAEs in each of structure-only CompStruct and CompStruct categories expressed in percentage with respect to the smaller MAE. The approximate number of training data (rounded to the nearest thousand) is recorded on the bottom right of each figure panel. The following are respectively the performance of structure-only CompStruct *vs.* CompStruct frameworks on non-poly *vs.* poly data segregation on the prediction of (C and J) band gap, $E_\mathrm{g}$ (D and K) Voigt-Reuss-Hill average bulk modulus, $K_\mathrm{VRH}$ (E and L) Voigt-Reuss-Hill average shear modulus, $G_\mathrm{VRH}$ (F and M) density, $\rho$, and (G and N) point density. Point density is a toy property (shaded in gray) solely dependent on structure, modified from density by substituting the atomic mass with the number of atoms in the unit cell in the density calculation.

## S3. A mathematical perspective on the prediction of properties with compositional information alone

In this section, we suggest a mathematical perspective on the interplay between the self-consistency between composition and structure and the choice of a dataset, and how this interplay enables the prediction of properties with only compositional information.

The self-consistency between composition and structure can be expressed as an implicit function $F(C, S) = 0$, where composition and structure are respectively abbreviated as *C*, and *S*. To illustrate the concept, we arbitrarily take this self-consistency function to be $C^2 + S^2 - 1 = 0$, namely a unit circle as shown in Figure S4A. In general, despite the self-consistency between (the implicit function of) *C* and *S*, *C* and *S* cannot be expressed by each other. However, the implicit function theorem [1] tells us when the input domain is limited, there exist regions such as the one marked in red in Figure S4B, where *S* can be expressed as a function of *C*, $S = \sqrt{1 - C^2}$, or $S = \sqrt{2} - C$ approximately. This expression of *S* with *C* signifies the embedding of structure in composition. Our data segregation in essence is to limit the input domain of the self-consistency function. Property, physically dictated to depend on both composition and structure, can be expressed as $P = g(C, S)$, where $g$ varies from property to property. If the input domain, namely the choice of the data segregation, is like Figure S4B, the property can be expressed by *C* alone due to the implicit function and the limited input domain, which translates to that property can be predicted from composition alone due to the self-consistency and choice of datasets.

However, in our stable data segregation, we don't observe comparably good prediction for every property, *e.g.*, density. We factor in property in Figure S4C. We argue the stable data segregation doesn't limit the input domain to a clean region where *S* can be completely expressed by *C*, like the red region in Figure S4C, but certain properties can still be expressed by *C* due to their property function $g$. For example, let $P = g(C, S) = C + S^2$. In this case, although *S* cannot be expressed by *C*, *P* can be expressed as $C + 1 - C^2$, where a sufficient amount of structure as $S^2$ (lacking sign information compared to the entirety of *S*) is embedded in composition *C*, allowing property to be independent of structure conditional on the selected data segregation/input domain. This can be the illustrative case for the stable data segregation with properties $E_f$, $E_g$, $K_{VRH}$, and $G_{VRH}$. For density, the property function might not effectively allow density to be effectively expressed by *C*, and thus for the poorer prediction of density with only compositional information on stable data segregation. (Admit the choice of $g(C, S) = C + S^2$ is non-optimal since it allows the property to be expressed by *C* for all input domain; still, it serves the purpose of conveying the concept.)

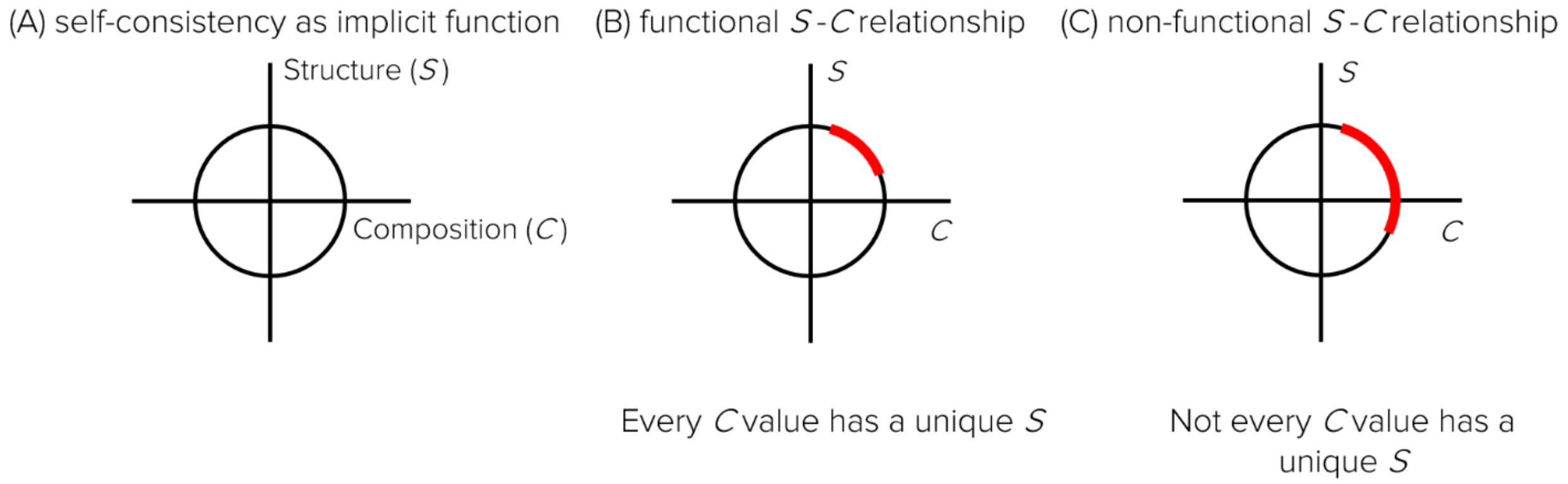

Figure S4. Illustration of self-consistency and choice of datasets. Structure and composition are respectively abbreviated as S and C in (B) and (C).

## S2. MAEs of Model Runs on various datasets

Table S1. Training MAE vs. test MAE of every model on every dataset.

| Model | Data Segregation | Property | Training MAE (averaged over nine runs) | Test MAE (averaged over nine runs) |
|---|---|---|---|---|
| Roost | Stable | $E_\mathrm{f}$ [eV/atom] | 0.017 | 0.028 |
| | | $E_\mathrm{g}$ [eV] | 0.172 | 0.341 |
| | | $K_\mathrm{VRH}$ [log(GPa)] | 0.024 | 0.087 |
| | | $G_\mathrm{VRH}$ [log(GPa)] | 0.022 | 0.119 |
| | | $\rho$ [g/cm$^3$] | 0.075 | 0.168 |
| | | point density [1e-3/Å$^3$] | 1.044 | 2.212 |
| | Unstable | $E_\mathrm{f}$ [eV/atom] | 0.057 | 0.157 |
| | | $E_\mathrm{g}$ [eV] | 0.202 | 0.424 |
| | | $K_\mathrm{VRH}$ [log(GPa)] | 0.045 | 0.162 |
| | | $G_\mathrm{VRH}$ [log(GPa)] | 0.052 | 0.247 |
| | | $\rho$ [g/cm$^3$] | 0.136 | 0.399 |
| | | point density [1e-3/Å$^3$] | 2.059 | 5.380 |
| CrabNet | Stable | $E_\mathrm{f}$ [eV/atom] | 0.009 | 0.027 |
| | | $E_\mathrm{g}$ [eV] | 0.069 | 0.293 |
| | | $K_\mathrm{VRH}$ [log(GPa)] | 0.025 | 0.078 |
| | | $G_\mathrm{VRH}$ [log(GPa)] | 0.018 | 0.104 |
| | | $\rho$ [g/cm$^3$] | 0.046 | 0.146 |
| | | point density [1e-3/Å$^3$] | 0.674 | 1.941 |
| | Unstable | $E_\mathrm{f}$ [eV/atom] | 0.036 | 0.149 |
| | | $E_\mathrm{g}$ [eV] | 0.101 | 0.384 |
| | | $K_\mathrm{VRH}$ [log(GPa)] | 0.083 | 0.158 |
| | | $G_\mathrm{VRH}$ [log(GPa)] | 0.115 | 0.238 |
| | | $\rho$ [g/cm$^3$] | 0.083 | 0.342 |
| | | point density [1e-3/Å$^3$] | 1.335 | 4.715 |
| MEGNet | Stable | $E_\mathrm{f}$ [eV/atom] | 0.012 | 0.024 |
| | | $E_\mathrm{g}$ [eV] | 0.172 | 0.298 |
| | | $K_\mathrm{VRH}$ [log(GPa)] | 0.013 | 0.073 |
| | | $G_\mathrm{VRH}$ [log(GPa)] | 0.015 | 0.092 |
| | | $\rho$ [g/cm$^3$] | 0.041 | 0.063 |
| | | point density [1e-3/Å$^3$] | 0.406 | 0.618 |
| | Unstable | $E_\mathrm{f}$ [eV/atom] | 0.056 | 0.105 |
| | | $E_\mathrm{g}$ [eV] | 0.197 | 0.392 |
| | | $K_\mathrm{VRH}$ [log(GPa)] | 0.049 | 0.184 |
| | | $G_\mathrm{VRH}$ [log(GPa)] | 0.044 | 0.267 |
| | | $\rho$ [g/cm$^3$] | 0.055 | 0.128 |
| | | point density [1e-3/Å$^3$] | 0.440 | 1.244 |
| CGCNN | Stable | $E_\mathrm{f}$ [eV/atom] | 0.015 | 0.022 |
| | | $E_\mathrm{g}$ [eV] | 0.063 | 0.260 |
| | | $K_\mathrm{VRH}$ [log(GPa)] | 0.010 | 0.072 |
| | | $G_\mathrm{VRH}$ [log(GPa)] | 0.011 | 0.092 |
| | | $\rho$ [g/cm$^3$] | 0.039 | 0.054 |
| | | point density [1e-3/Å$^3$] | 0.378 | 0.630 |
| | Unstable | $E_\mathrm{f}$ [eV/atom] | 0.030 | 0.087 |
| | | $E_\mathrm{g}$ [eV] | 0.095 | 0.345 |
| | | $K_\mathrm{VRH}$ [log(GPa)] | 0.032 | 0.141 |
| | | $G_\mathrm{VRH}$ [log(GPa)] | 0.053 | 0.194 |
| | | $\rho$ [g/cm$^3$] | 0.038 | 0.086 |
| | | point density [1e-3/Å$^3$] | 0.444 | 1.112 |